\documentclass[
    ,final            
  ]
  {aipproc}

\layoutstyle{6x9}
\usepackage{amsmath}
\usepackage{array}
\usepackage{float}	
\usepackage{amssymb}
\usepackage{color}

\begin{document}

\title{Suprathermal viscosity of dense matter}
 

\author{Mark Alford}{
  address={Department of Physics, Washington University in St. Louis, Missouri, 63130, USA}
}

\author{Simin Mahmoodifar}{
  address={Department of Physics, Washington University in St. Louis, Missouri, 63130, USA}
}

\author{Kai Schwenzer}{
  address={Department of Physics, Washington University in St. Louis, Missouri, 63130, USA}
}

\begin{abstract}
Motivated by the existence of unstable modes of compact stars that eventually grow large, we study the bulk viscosity of dense matter, taking into account non-linear effects arising in the large amplitude regime, where the deviation $\mu_\Delta$ of the chemical potentials 
from chemical equilibrium fulfills $\mu_\Delta \gtrsim T$. 
We find that this supra-thermal bulk viscosity can provide a potential mechanism for saturating unstable modes in compact stars since the viscosity is strongly enhanced.
Our study confirms previous results on strange quark matter and shows that the suprathermal enhancement is even stronger in the case of hadronic matter. We also comment on the competition of different weak channels and the presence of suprathermal effects in various color superconducting phases of dense quark matter.
  \end{abstract}

\maketitle


\section{Introduction}

Bulk viscosity is an important mechanism for the damping of oscillations of compact stars \cite{Papaloizou:1978,Andersson:1997xt,Friedman:1997uh,Lindblom:1999yk,Madsen:1999ci} and therefore is a useful measure for differentiating possible phases of dense matter in their interior. 
At sufficiently low viscosity and high rotation rate, r-modes
are unstable and can cause rapid spin-down of the star via gravitational
radiation \cite{Owen:1998xg} if their growth is stopped by some non-linear mechanism.  Finding the relevant mechanism is important because it determines the amplitude at which the r-mode saturates, and hence the rate at which it spins down the star.
At low amplitudes the bulk viscosity is amplitude-independent, but
since the r-mode is unstable its amplitude grows, and unless 
stopped by other mechanisms \cite{Lindblom:2000az,Gressman:2002zy,Lin:2004wx,Bondarescu:2007jw} will quickly enter the
``supra-thermal'' regime where the bulk viscosity grows with amplitude \cite{Madsen:1992sx,Reisenegger:2003pd,Bonacic:2003th},
and may become large enough to stop the growth of the mode. 

This proceeding is based on our recent paper \cite{Alford:2010gw}, where we discuss the large amplitude behavior of the bulk viscosity in detail. It expands the analytic approximation methods and extends the analysis to other phases of dense matter.

\section{General formalism for the bulk viscosity}

The bulk viscosity of a given form of matter is defined by the
response of the system to an oscillating compression and rarefaction. 
This corresponds to an oscillation in the densities of all
exactly conserved quantities.
In compact stars the relevant quantity is typically the baryon number density,
\begin{equation}
n_{*}(\vec r,t) = \bar{n}_{*}(\vec r)+\delta n_{*}(\vec r,t)
 =\bar{n}_{*}(\vec r)+\Delta\! n_{*}(\vec{r})\sin\left(\omega t\right)\,,
\end{equation}
where $\bar n_{*}$ is the equilibrium density and $\Delta n_*$ is the amplitude of the oscillation. We assume $\Delta n_{*}\ll\bar{n}_{*}$. 
The bulk viscosity is then given by  \cite{Madsen:1992sx}
\begin{equation}
\zeta\approx \frac{2}{\omega^{2}}\left\langle \frac{d\epsilon}{dt}\right\rangle_{\rm \! diss} \frac{\bar{n}_{*}^{2}}{\left(\Delta\! n_{*}\right)^{2}}\,.
\label{eq:zetadef}
\end{equation}
Where $\left\langle d\epsilon / dt\right\rangle_{\rm \! diss} $ is the averaged energy dissipation rate per volume in the fluid due to the oscillation.

Due to the driving density fluctuation the quantity $\mu_\Delta \equiv \sum_i \mu_i - \sum_f \mu_f $ is driven out of equilibrium and its re-equilibration
leads to the bulk viscosity. We will assume that the latter
arises from beta-equilibration of fermionic
species and that the net rate takes the general form
\begin{equation}
\Gamma^{(\leftrightarrow)} = -\tilde{\Gamma} T^{\kappa}\mu_\Delta \left(1+\sum_{j=1}^{N}\chi_j\left(\frac{\mu_\Delta^{2}}{T^{2}}\right)^{j}\right)\,.
\label{eq:gamma-parametrization}
\end{equation}
where $N$ determines the highest power of $\mu_\Delta$
arising in the rate. The sum merely reflects that phase space for the weak processes is either provided by softening the Fermi surface due to finite temperature or by shifting the chemical potentials from their equilibrium values, so that $\kappa=2N$.
In terms of dimensionless variables
$\varphi\equiv\omega t$ and ${\cal A} \left(\varphi\right)\equiv\frac{\mu_\Delta\left(t\right)}{T}$,
the chemical fluctuation then fulfills the differential equation,

\begin{equation}
\frac{d {\cal A}}{d\varphi}=d \cos\left(\varphi\right)-f {\cal A} \left(1+\sum_{j=1}^{N}\chi_{j}{\cal A}^{2j}\right)\,,
\label{eq:general-diff-eq}
\end{equation}
with the prefactors of the driving term and feedback term given by
\begin{equation}
d\equiv\frac{C}{T}\frac{\Delta n_{*}}{\bar{n}_{*}}\quad,\quad f\equiv\frac{B\tilde{\Gamma}T^{\kappa}}{\omega}\,.
\label{eq:general-parameters}
\end{equation}
Where B and C are susceptibilities that can be obtained from the equation of state of matter.
Note that the feedback term involves both linear and non-linear parts which are controlled by a single parameter $f$ and that its particular form is determined by the constants $\chi_j$ which parametrize the particular weak rate. The viscosity is then given by
\begin{equation}
\zeta=-\frac{\tilde{\Gamma} C T^{\kappa+1}}{\pi \omega^2}\frac{\bar{n}_{*}}{\Delta\! n_{*}}  \int_{0}^{2\pi}\int_{0}^{\varphi} {\cal A(\varphi^\prime)} \left( 1+\sum_{j=1}^{N}\chi_j \left({\cal A}(\varphi^\prime)\right)^{j} \right) d\varphi^{\prime}\cos\left(\varphi\right)d\varphi\,. \label{eq:general-solution2}
\end{equation}
where ${\cal A}\left(\varphi;d,f\right)$ is the periodic solution to eq.~(\ref{eq:general-diff-eq}) (for details see \cite{Alford:2010gw}). 


In the sub-thermal limit  where $\mu_\Delta \ll T$, corresponding to ${\cal A} \ll1$, the non-linear
terms in eq.~(\ref{eq:general-diff-eq}) can be neglected, leading to the following equation for the sub-thermal bulk viscosity
\begin{equation}
\zeta^{<}=\frac{C^{2}\tilde{\Gamma}T^{\kappa}}{\omega^{2}+(B \tilde{\Gamma} T^{\kappa})^2}
= \zeta_{max}^< \frac{2 \omega B\tilde\Gamma T^\kappa}{\omega^2 
+ (B\tilde\Gamma T^\kappa)^2}
\,.
\label{eq:sub-viscosity}
\end{equation}
As can be seen from this equation, the sub-thermal bulk viscosity is amplitude-independent. 
As long as the combination of susceptibilities $C^2/B$ does not
vary too quickly with temperature, the sub-thermal viscosity has a maximum
\begin{equation}
\zeta_{max}^{<}=\frac{C^{2}}{2\omega B}\quad\mathrm{at}\quad T_{max}=\left(\frac{\omega}{\tilde{\Gamma}B}\right)^{\frac{1}{\kappa}}\,.
\label{eq:sub-maximum}
\end{equation}
At sufficiently low temperatures and high frequency one has $f \ll 1$, and therefore the feedback terms in eq.~(\ref{eq:general-diff-eq}) are negligible and the viscosity in this intermediate regime can be evaluated as
\begin{equation}
\label{eq:madsen-approximation}
\zeta^\sim = \frac{C^{2}\tilde{\Gamma}T^{\kappa}}{\omega^{2}}\left(1+\sum_{j=1}^{N}\frac{\left(2j+1\right)!! \chi_{j}}{2^{j}\left(j+1\right)!}\left(\frac{C}{T} \frac{\Delta n}{\bar{n}}\right)^{2j}\right)
\end{equation}
which can be combined with the subthermal result for an analytic solution in both regions $\zeta^\lesssim=\zeta^<+\theta(T-T_{max}) \zeta^\sim$.

In the supra-thermal limit where $\mu_\Delta\gg T$, corresponding to ${\cal A} \gg1$,
 only the largest power of ${\cal A}$ is relevant in
eq.~(\ref{eq:general-diff-eq}) 
and therefore the viscosity scales in this limit as
\begin{equation}
\zeta\sim\left(\frac{\Delta n_{*}}{\bar{n}_{*}}\right)^{-\frac{2N}{2N+1}}\label{eq:supra-limit}
\end{equation}
and decreases again at very large amplitudes.

\begin{figure*}
\begin{minipage}[t]{0.5\textwidth}%
\includegraphics[width=\hsize]{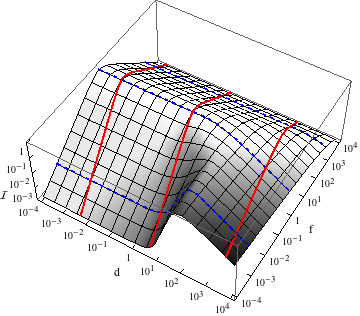}%
\end{minipage}%
\begin{minipage}[t]{0.5\textwidth}%
\hspace*{0.05\hsize} \includegraphics[width=0.9 \hsize]{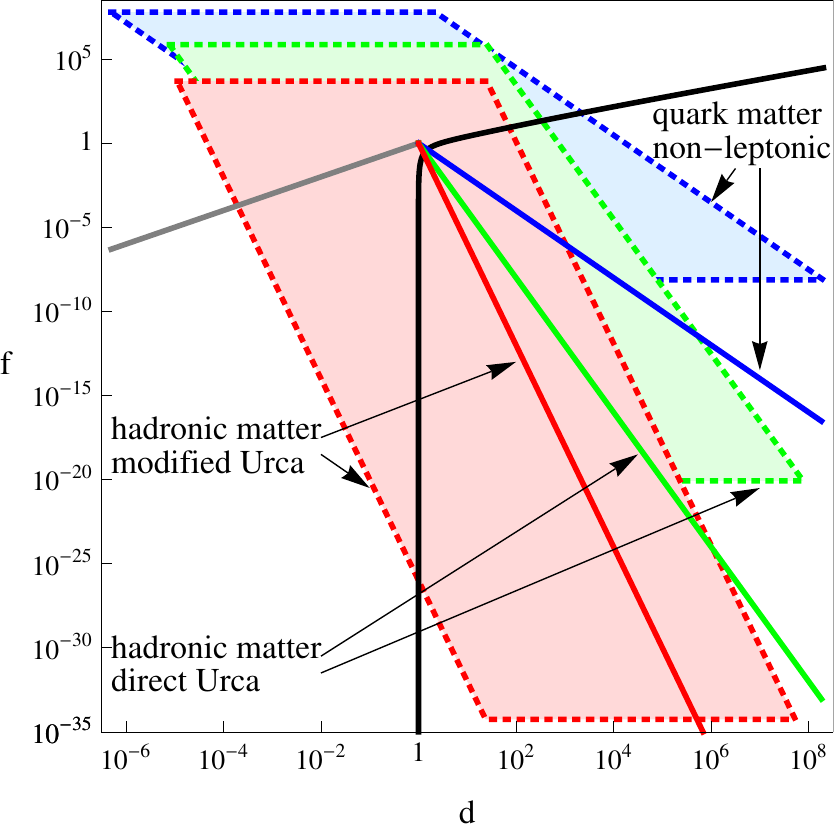}%
\end{minipage}
\caption{\label{fig:gen-sol} {\em Left panel:}
The function ${\cal I}$ arising in the general
solution eq.~(\ref{eq:general-parametrization}). The plot is given for the case of hadronic matter with modified Urca equilibration but the generic qualitative form is independent of the considered form of matter. 
The function has a global maximum of $1$
reached asymptotically for $d\to 0,f=1$ and a line of slowly decreasing
local maxima along a parabola in the $d$-$f$ plane. The shading
of the surface denotes the size of the amplitude ${\cal A}$ so that
dark shades of grey represent the supra-thermal regime. 
Eq.~(\ref{eq:general-parameters}) relates $d$ and $f$ to underlying
physical parameters such as temperature $T$ and amplitude.
An amplitude increase (keeping all other variables fixed) results in a linear increase in the variable
$d$ as shown by the dashed curves. An increase in temperature
changes the viscosity along a line
shown by the solid curves. {\em Right panel:} The relevant ranges of the parameters $d$ and $f$ in the general parametrization eq. (\ref{eq:general-parametrization}) when the underlying parameters are varied in the physically interesting ranges $T\in [10^5,10^{11}]$ K, $\Delta n/\bar n \in [10^{-5},1]$, $\omega\in [1,1000]$ Hz, $\bar n\in [1/4,10] \, n_0$ are denoted by the the dotted boundaries for hadronic matter with modified and direct Urca reactions and strange quark matter with non-leptonic reactions. The analytic sub-thermal approximation eq. (\ref{eq:sub-viscosity}) is valid well to the left of the black line, whereas the intermediate linear approximation eq. (\ref{eq:madsen-approximation}) is valid well below the piecewise curve formed by the gray linear segment on the left and the corresponding linear segment for the respective form of matter.}
\end{figure*}

The general result for the bulk viscosity can be written in the following form

\begin{equation}
\zeta =\zeta^<_{max} \, {\cal I} \left(d,f\right)= \frac{C^2}{2\omega B}{\cal I} \left(d,f\right) \label{eq:general-parametrization}
\end{equation}
where the dimensionless function ${\cal I}$ that includes the
non-trivial parameter dependence is given by
\begin{equation}
{\cal I}(d,f)\equiv\frac{2}{\pi d}\int_{0}^{2\pi} {\cal A}(\varphi;d,f)\cos(\varphi)d\varphi \,
\end{equation}
The generic form of the function ${\cal I}(d,f)$ is shown in the left panel of Figure~\ref{fig:gen-sol}
It has a global maximum value of 1, reached in the sub-thermal limit and a line of local maxima along a parabola in the $d$-$f$-plane. Thus the maximum value (\ref{eq:sub-maximum}) of the sub-thermal viscosity is also the maximum in the general case and depends only
on the equation of state, the density and the frequency but is independent
of the weak rate. The weak rate influences, however, at what temperatures
and amplitudes the local maxima are reached. On the right panel of figure~\ref{fig:gen-sol}, the physically interesting regions of the parameters $d$ and $f$ for different forms of matter discussed below are compared to the regions of validity of the analytic approximations discussed above. Although the analytic approximations cover large parts of the parameter space, non-linear effects nevertheless require a numeric solution for extreme parameter values.

\section{Bulk viscosity for different types of dense matter}

In this section we apply our results given above to various models of quark matter and hadronic matter. We study strange quark matter \cite{Witten:1984rs} both within a simple quark gas model of free quarks in a "confining bag" as well as a phenomenological parameterization of the quark matter equation of state proposed in \cite{Alford:2004pf} where a new parameter $c$ has been introduced which incorporates some effects of strong interactions between the quarks.
For hadronic matter we consider in addition to a free hadron gas
the well-known model by Akmal, Pandharipande
and Ravenhall \cite{Akmal:1998cf} which relies on a 
potential model that reproduces scattering data at nuclear densities.
As a low density extension of the APR data we use \cite{Baym:1971pw,Negele:1971vb}. 
In order to 
to apply our general results to other equations
of state, away from chemical equilibrium we use the simple quadratic parameterization in terms of the symmetry energy employed in \cite{Lattimer:1991ib}.
The resulting strong interaction parameters describing the response of the different models are given in Table~\ref{tab:strong-parameters}.

\begin{table}[htb]
\newcommand{\st}{\rule[-3.5ex]{0em}{8.5ex}}  
\begin{tabular}{l@{\quad}c@{\quad}c@{\quad}c}
\hline 
\rule[-2ex]{0em}{5ex} & & $B$ & $C$\tabularnewline
\hline 
\st quark matter (quark gas: c=0) & & $\frac{2\pi^{2}}{3(1-c)\mu_{q}^{2}}\left(1\!+\!\frac{m_{s}^{2}}{12(1-c)\mu_{q}^{2}}\right)$ & $-\frac{m_{s}^{2}}{3(1-c)\mu_{q}}$\tabularnewline
\hline 
\st hadronic matter & & $\frac{8S}{n}\!+\negthinspace\frac{\pi^{2}}{\left(4\left(1\!-\!2x\right)S\right)^{2}}$ & $4\!\left(1\!-\!2x\right)\!\left(\! n\!\frac{\partial S}{\partial n}\!-\!\frac{S}{3}\!\right)$\tabularnewline
\hline
\st free hadron gas & & $\frac{4m_{N}^{2}}{3\left(3\pi^{2}\right)^{\frac{1}{3}}n^{\frac{4}{3}}}$ & $\frac{\left(3\pi^{2}n\right)^{\frac{2}{3}}}{6m_{N}}$\tabularnewline
\hline
\end{tabular}
\caption{\label{tab:strong-parameters}
Susceptibilities B and C for different model equations of state. In the case of 
hadronic matter with baryon density $n$ a quadratic ansatz in the proton fraction $x$ parameterized by the symmetry energy $S$ is employed. The expressions for a free hadron
gas are given to leading order in $n/m_{N}^{3}$, and for quark matter with quark chemical potential $\mu_q$ to next to leading order in $m_{s}/\mu_q$. The parameter $c$ takes into account interaction effects within the employed quark matter model and vanishes for an ideal quark gas.
}
\end{table}

In unpaired strange quark matter, consisting of $u$, $d$ and $s$ quarks, the dominant channel for beta equilibration is the non-leptonic flavor changing process, $d+u\leftrightarrow s+u$,
where the equilibrating chemical potential in response to the the driving oscillation is given by $\mu_\Delta=\mu_{s}-\mu_{d}$.
In the case of hadronic matter we assume that weak equilibration occurs via the Urca channel
\begin{equation}
p+e^{-}\to n+\nu_{e}\quad,\quad n\to p+e^{-}+\bar{\nu}_{e}\end{equation}
There are two qualitatively different cases depending on whether the
direct process is possible or only the modified version where a bystander nucleon is necessary to satisfy energy-momentum conservation. 
Here the driving baryon number density oscillation yields the oscillating
chemical potential difference $\mu_\Delta=\mu_{n}-\mu_{p}-\mu_{e} \ $.
The equilibration rate parameters $\tilde{\Gamma}$, $\kappa$ and $\chi_{i}$, defined in
eq.~(\ref{eq:gamma-parametrization}), for the weak processes discussed above, are given in Table~\ref{tab:weak-parameters}.

\begin{table}[htb]
\newcommand{\st}{\rule[-3ex]{0em}{8ex}}  
\begin{tabular}{l@{\quad}c@{\quad}c@{\quad}c@{\quad}c@{\quad}c}
\hline 
\rule[-1.5ex]{0em}{4ex} Matter/Channel
  & $\tilde{\Gamma} \,\bigl[{\rm MeV}^{(3-\kappa)}\bigr]$ 
  & $\kappa$ & $\chi_{1}$ & $\chi_{2}$ & $\chi_{3}$\tabularnewline
\hline 
\st quark non-leptonic & $6.59\!\times\!10^{-12}\, \Bigl(\frac{\mu_{q}}{300~{\rm MeV}}\Bigr)^{5}$ 
  & $2$ & $\frac{1}{4\pi^{2}}$ & $0$ & $0$\tabularnewline
\hline 
\st hadronic direct Urca & $5.24\!\cdot\!10^{-15} \left(\!\frac{x\, n}{n_{0}}\!\right)^{\!\frac{1}{3}}$ 
  & $4$ & $\frac{10}{17\pi^{2}}$ & $\frac{1}{17\pi^{4}}$ & $0$\tabularnewline
\hline 
\st hadronic modified Urca & $4.68\!\cdot\!10^{-19} \left(\!\frac{x\, n}{n_{0}}\!\right)^{\!\frac{1}{3}}$ 
  & $6$ & $\frac{189}{367\pi^{2}}$ & $\frac{21}{367\pi^{4}}$ & $\frac{3}{1835\pi^{6}}$\tabularnewline
\hline
\end{tabular}
\caption{\label{tab:weak-parameters}Weak interaction parameters describing
the considered damping process. Here $\mu_q$ is the quark chemical potential, $n$ is the baryon density, $n_0$ nuclear saturation density and $x$ the proton fraction.
}
\end{table}

\begin{figure*}
\begin{minipage}[t]{0.5\textwidth}%
\includegraphics[scale=.65]{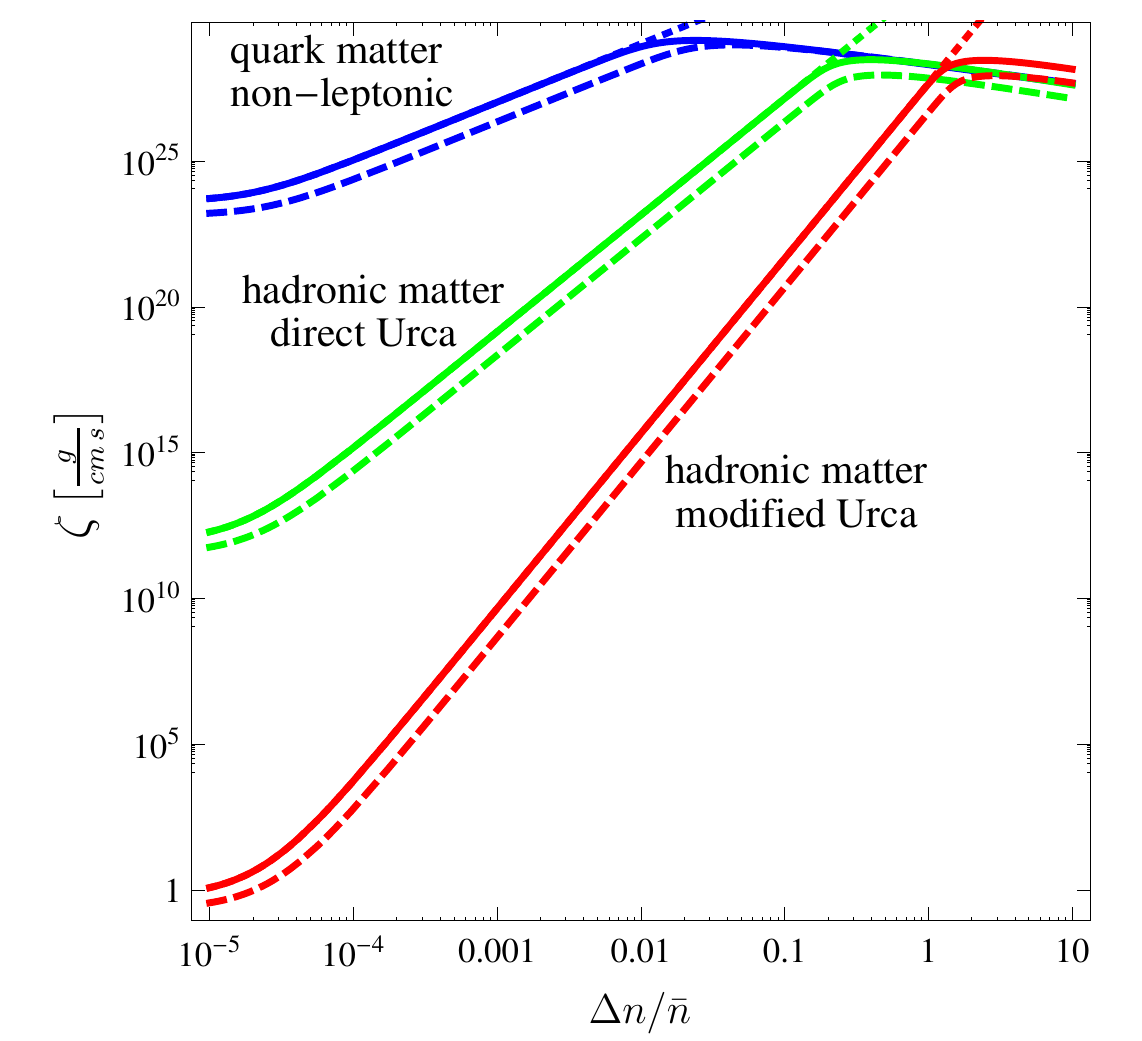}%
\end{minipage}%
\begin{minipage}[t]{0.5\textwidth}%
\includegraphics[scale=.65]{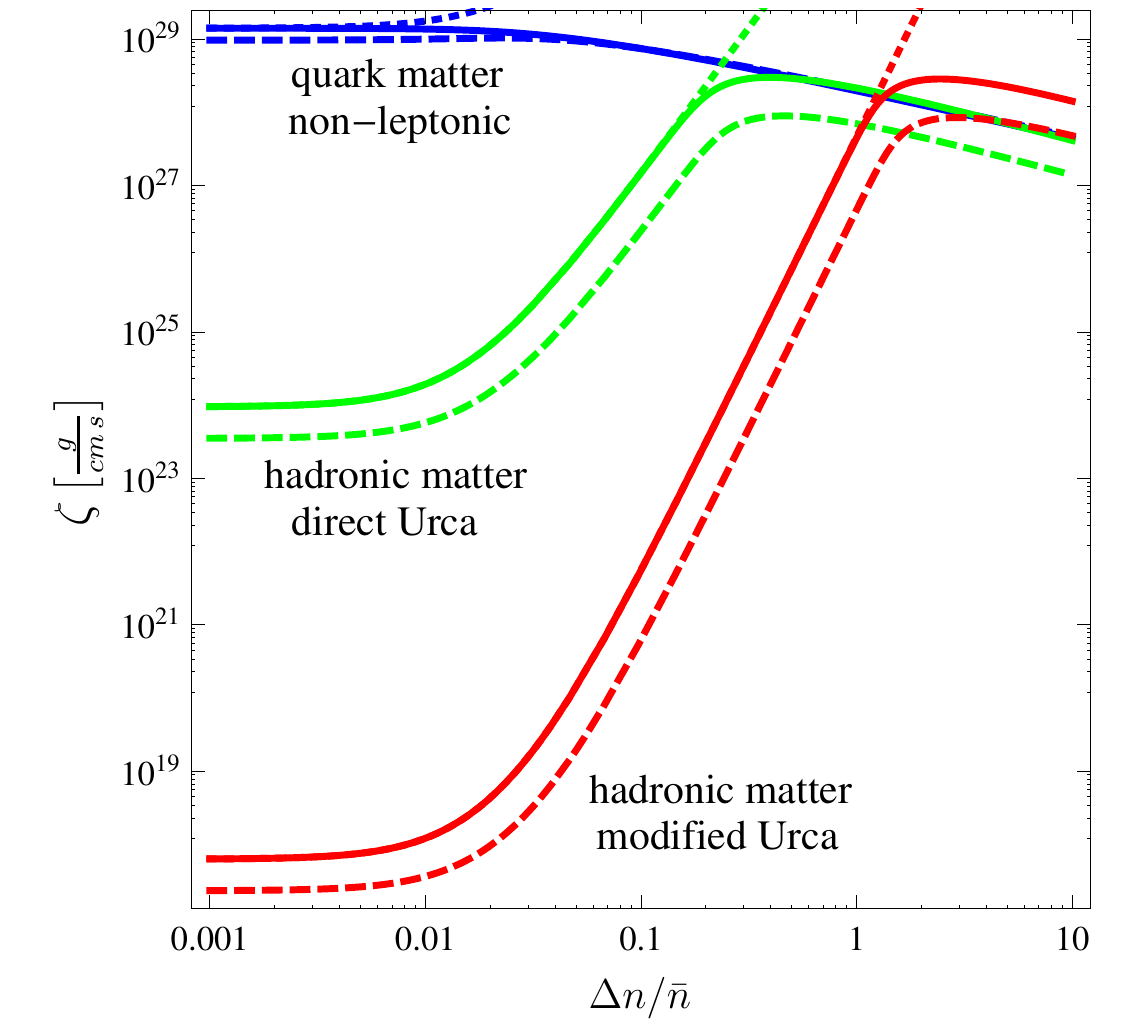}%
\end{minipage}

\caption{\label{fig:all-viscosity}
Comparison of the bulk viscosity of the different
forms of matter studied in this work as a function of the
density oscillation amplitude $\Delta n/\bar{n}$. The frequency is
$\omega=8.4$~kHz, corresponding to an r-mode in a millisecond pulsar and $\bar n = 2n_0$.
{\em Left panel:}
low temperature $T=10^{6}$~K; {\em right panel:} high temperature $T=10^{9}$~K.
The dashed curves are for the free hadron
and free quark models; the solid curves are for APR hadron matter,
and interacting quarks with $m_s=150$\,MeV and 
$c=0.3$ .
}
\end{figure*}

Using eqs.~(\ref{eq:sub-viscosity}-\ref{eq:general-parametrization}) and the parameters given in Table~\ref{tab:strong-parameters} and \ref{tab:weak-parameters}, we obtain the results shown in Fig.~\ref{fig:all-viscosity},  where the bulk viscosity as a function of the density oscillation amplitude at two temperatures 
is given. The solid lines show the results for interacting matter whereas the dashed lines show the
free hadron/quark gas results. The dotted lines, 
which are in most places invisible beneath the solid lines,
show the combined analytic approximation below eq.~(\ref{eq:madsen-approximation}) valid at moderate amplitudes. 
At the lower temperature the viscosity reaches
the supra-thermal regime, already for small amplitudes, whereas 
at the higher temperature the
sub-thermal regime, in which the viscosity is independent of the amplitude, extends to large amplitudes.
The stronger non-linear feedback
in the hadronic cases leads to a significantly steeper rise that correlates
with the largest power in eq.~(\ref{eq:gamma-parametrization}). Interestingly,
despite these differences the maximum value reached by varying the
amplitude is still roughly the same as the maximum value in the sub-thermal
limit eq.~(\ref{eq:sub-maximum}). This means
that oscillations are approximately equally damped at all temperatures once the amplitude becomes sufficiently large. The maximum arises for amplitudes
of the order $0.01$, $0.1$ and $1$ for strange quark matter and
hadronic matter with direct and modified Urca, respectively. Therefore 
the simple analytic expression eq.~(\ref{eq:madsen-approximation})
provides a remarkably good approximation for modified Urca processes.
The supra-thermal enhancement of the bulk viscosity is
so strong
that it could well provide the main saturation
mechanism for unstable r-modes, stopping their growth at amplitudes
that are below the threshold for other competing saturation mechanisms
but large enough to allow 
spin-down of a neutron star
via gravitational radiation on astrophysical time scales.

The discussed suprathermal effects should be important for the competition of different weak channels contributing to the viscosity. This has been considered for the non-leptonic and Urca channels in strange quark matter in \cite{Sa'd:2007ud}. From the stronger temperature dependence of the corresponding quark Urca rates it is clear that analogous to the direct hadronic Urca case the semi-leptonic contributions to the viscosity also have a stronger amplitude dependence than the non-leptonic contribution. Therefore, at large amplitudes the mixing between these two channels should be far more pronounced than what has been observed previously in the subthermal case \cite{Sa'd:2007ud}. Similar conclusions hold for the mixing in hadronic matter when the hyperonic channel is included \cite{Jones:2001ya}.

Our general results for the bulk viscosity can also be applied to other forms of dense matter. For example in 2SC color-superconducting quark matter, where all strange quarks as well as up and down quarks of one color remain unpaired \cite{Alford:2007xm}, it has been shown that at $T\ll T_c$ the rate of the non-leptonic flavor changing process $d+u\leftrightarrow s+u$ differs from that in the unpaired quark matter only by a constant factor, $\Gamma^{2SC}=\Gamma^{unpaired}/9$ \cite{Alford:2006gy}. The susceptibilities B and C receive additional contributions involving the gap parameter $\Delta$, but since $\Delta<m_s$ there is not a significant change and one can expect the same qualitative behavior for the bulk viscosity of the 2SC quark matter as in unpaired quark matter. According to eq.~(\ref{eq:general-parametrization}) the maximum value of the bulk viscosity $\zeta_{max}$ is independent of the rate and thereby only slightly changed, but the bulk viscosity of the 2SC quark matter reaches its maximum value at higher amplitudes compared to unpaired quark matter. 
A similar statement holds for the bulk viscosity of 
1SC phases \cite{Sa'd:2006qv,Wang:2010yd}. Here the corresponding suppression factor depends on the particular pairing pattern, which may involve only part of the quarks on the Fermi surface, and can be as low as $0.03$, thereby enhancing the effect discussed for 2SC pairing.

In color-flavor locked (CFL) matter \cite{Alford:2007xm} all quark species are gapped and the rates for processes involving either quarks or massive pseudo Goldstone bosons, as well as the corresponding viscosities, are exponentially suppressed below the critical temperature of the order of tens of MeV \cite{Alford:2007rw}. The only ungapped modes are the superfluid phonons, but since their rate is not given by weak interactions \cite{Manuel:2007pz} a simple estimate employing our general expressions is not possible in this case. Yet, we expect that neither the suppressed modes nor the phonons are very sensitive to changes in the underlying quark chemical potentials. Correspondingly, the amplitude dependence of the CFL viscosity should only be moderate and should not qualitatively change the picture obtained in \cite{Alford:2007rw,Manuel:2007pz}. In contrast, in the kaon condensed CFLK0 phase there are truly massless Goldstone bosons that mediate weak interactions \cite{Alford:2008pb}. Even though these particles are bosonic their weak reactions should be similarly influenced by changes in the corresponding flavor chemical potentials and therefore we expect a suprathermal enhancement at large amplitudes.
  
\section{Conclusion}
In this paper we have studied the bulk viscosity of dense matter taking into account non-linear effects that arise in the supra-thermal regime. We gave an approximate analytic solution as well as a general numerical solution for the bulk viscosity which are valid for all types of matter where equilibration occurs via fermions. 
We confirm previous results for the amplitude dependence
of the bulk viscosity of strange quark matter \cite{Madsen:1992sx}
and find that these supra-thermal effects are parametrically even more important
in nuclear matter.
Whereas our result directly extend to partly gapped phases like 1SC and 2SC, in the fully gapped CFL phase the suprathermal enhancement should be mild.
The most obvious application of our results is to the damping
of unstable r-mode oscillations in neutron stars. As the amplitude
of the mode enters the supra-thermal regime
the viscosity will increase steeply above the
sub-thermal result and can exceed it by many orders of magnitude. It could therefore provide an adequate saturation mechanism if it occurs before other non-linear dynamic
effects \cite{Lindblom:2000az,Gressman:2002zy,Lin:2004wx,Bondarescu:2007jw} start to operate.


\begin{theacknowledgments}
M.G.A. thanks the organizers of the QCD@Work Guiseppe Nardulli memorial workshop for arranging a very successful conference.
This research was supported in part by the
Offices of Nuclear Physics and High Energy Physics of the
U.S. Department of Energy under contracts
\#DE-FG02-91ER40628,  
\#DE-FG02-05ER41375. 
\end{theacknowledgments}

\newcommand{\apjl}{Astrophys. J. Lett.\ }
\newcommand{\mnras}{Mon. Not. R. Astron. Soc.\ }
\newcommand{\aap}{Astron. Astrophys.\ }

\bibliographystyle{aipproc}   


\bibliography{BVproceeding}

\IfFileExists{\jobname.bbl}{}
 {\typeout{}
  \typeout{******************************************}
  \typeout{** Please run "bibtex \jobname" to optain}
  \typeout{** the bibliography and then re-run LaTeX}
  \typeout{** twice to fix the references!}
  \typeout{******************************************}
  \typeout{}
 }

\end{document}